\documentclass{webofc}

\usepackage[varg]{txfonts}   
\usepackage{hyperref}
\usepackage{url}
\usepackage{mathptmx}
\usepackage[T1]{fontenc}
\usepackage[utf8]{inputenc}
\usepackage{graphicx}
\usepackage{amsmath}
\usepackage{booktabs}
\usepackage{xcolor}
\usepackage{caption}
\RequirePackage{orcidlink}
\hypersetup{colorlinks=true,citecolor=blue,urlcolor=blue,linkcolor=blue}
\newcommand{\code}[1]{\texttt{\small #1}}

\usepackage{geometry}
\begin{document}
\title{CMS Open Data Visualization with FireworksWeb}
%
%

\author{
\firstname{Yuxiao} \lastname{Wang}\inst{1}\orcidlink{0009-0004-1228-9849}\fnsep\thanks{\email{yuxiao.wang@cern.ch}}
\and
\firstname{Alja} \lastname{Mrak-Tadel}\inst{2}\orcidlink{0000-0002-3392-8345}
\and
\firstname{Matev\v{z}} \lastname{Tadel}\inst{2}\orcidlink{0000-0001-8800-0045}
\and
\firstname{Avi} \lastname{Yagil}\inst{2}\orcidlink{0000-0002-6108-4004}
\and
\firstname{Mario} \lastname{Masciovecchio}\inst{2}\orcidlink{0000-0002-8200-9425}
\and
\firstname{Dmytro} \lastname{Kovalskyi}\inst{3}\orcidlink{0000-0002-6923-293X}
\and
\firstname{Sergey} \lastname{Linev}\inst{4}\orcidlink{0000-0003-4711-9327}
}

\institute{Tsinghua University, Beijing, China 
\and
    University of California San Diego, La Jolla, CA, USA
\and
    Massachusetts Institute of Technology, Cambridge, MA, USA
\and
    GSI Helmholtzzentrum f\"ur Schwerionenforschung, Darmstadt, Germany
          }

\abstract{FireworksWeb is a web-based event display utilizing a C++ ROOT-Eve backend with SAPUI5 frontend for interactive 3D visualization of particle physics events directly in the browser. Building upon ROOT-Eve and RenderCore, it eliminates local software installation while maintaining professional-grade event display capabilities. FireworksWeb is currently deployed for live event monitoring in the CMS control room, demonstrating its reliability for real-time data visualization in production environments.

The CMS experiment has made a large amount of collision datasets publicly available through the CERN Open Data portal, which are the valuable resources for education, outreach, and research. We have extended FireworksWeb to enable visualization of these open datasets, making particle physics data accessible to diverse audiences without requiring local software installation.

Our prototype implementation includes standalone cameras, event filters, configurable collection settings, and projection controls with fish-eye distortion for cylindrical geometries. The web architecture communicating via CGI enables dataset queries and filtered event selection based on physics criteria. We demonstrate the prototype with CMS Open Data events, discuss the technical architecture, and outline applications for education, public outreach, and exploratory physics analysis.
}
\maketitle
\thispagestyle{plain}

\section{Introduction}
\label{intro}

Event displays are a standard tool in collider physics. They turn the reconstructed content of an event (tracks, calorimeter deposits, muons, jets, and vertices) into a picture that a physicist can inspect directly, and they are useful for teaching and outreach as well as for debugging reconstruction. In CMS this role has long been filled by Fireworks~\cite{fireworks2010, cmsevd2011}, a physics-analysis display built on the ROOT Event Visualization Environment (ROOT-Eve or REve for short)~\cite{root,eve}. Fireworks shows reconstructed, high-level objects against an idealized geometry, uses a schematic representation rather than a geometrically exact one, and is distributed as a self-contained binary.
 
This way of delivering the tool has become harder to maintain. The desktop application depends on OpenGL, whose support is deprecated on macOS and awkward over remote connections. It needs platform-specific builds and a local CMSSW-Light/ROOT installation, and it serves a single user, with no simple path to remote or shared use. At the same time, the CMS Open Data programme~\cite{opendata} has released large samples of real collision data to the public. These data are useful for education, outreach, and research, but the raw ROOT Event Data Model files cannot be read without dedicated tooling. Our aim is that anyone with a browser can look at real CMS collision events, with nothing to
install.
 
FireworksWeb addresses this. It reimplements Fireworks as a C\texttt{++} server driving a thin browser client, built on REve, the event-display module of ROOT~7. Two design choices carry over from Fireworks. First, the display is physics-oriented: it shows reconstructed objects against an idealized geometry and prefers a schematic drawing where that reads more clearly. Second, the same
tool serves two audiences. For the physicist it has to stay precise, able to filter collections, inspect a single hit, and reproduce a view. For the student or visitor it has to work from a shared link with no instructions. Keeping both in one application, rather than two, motivates several of the features below. The rest of the paper covers the architecture and rendering path, the camera and projection controls, view persistence, the Open Data pipeline, the universal display and its VSDNano toolkit, and the plans for the HL-LHC.

\section{From Fireworks to FireworksWeb}
The move from the desktop program to the current web application spans more than fifteen years. Fireworks began in 2008 as a ROOT TEve OpenGL-based desktop program. Work on a web version started in 2018 with a study of the server/client model~\cite{exploring2019}, and in 2019 an REve prototype showed physics collections in the browser~\cite{eve7fw2020}. By 2021 the application was deployed at CERN and UCSD and used in the CMS control room. In 2022 the 3D
rendering was moved from THREE.js to RenderCore. The 2024--25 period brought the toolkit to Run~3 production use, adding simulated-data visualization, CaloTower handling, and dockable views, and the work reported here (2025--26) extends it to the Open Data.
 
The Run~3 cycle touched every layer. For data and collections, the server gained support for the CMSSW\_14\_2 formats, for simulated data (SimData), and for the muon RecHits of the GEM, ME0, RPC, and DT subdetectors. For views and interaction, views can be docked and undocked, a geometry browser was added, event labels and axes can be drawn, CaloTower corrections were applied, and projection compression was reworked. For rendering and infrastructure, RenderCore brought faster geometry drawing and precise mouse picking, and per-collection controllers and better error reporting made the application steadier for control-room use.
 
Some context on the lineage. TEve was the original OpenGL desktop visualization environment of ROOT~\cite{eve}. REve is its reimplementation for ROOT~7, using ROOT's HTTP server to talk to a browser client~\cite{eve7fw2020}; REve is the resulting event-display module, and FireworksWeb is the CMS application on top of it. The early prototypes used JSROOT and THREE.js for rendering. The move to
RenderCore (Sec.~\ref{sec:rendercore}) is what made rendering of large data-collections and geometries practical in the browser.

\section{Architecture}
\label{sec:arch}
\begin{figure}[t]
\centering
\includegraphics[width=0.7\linewidth]{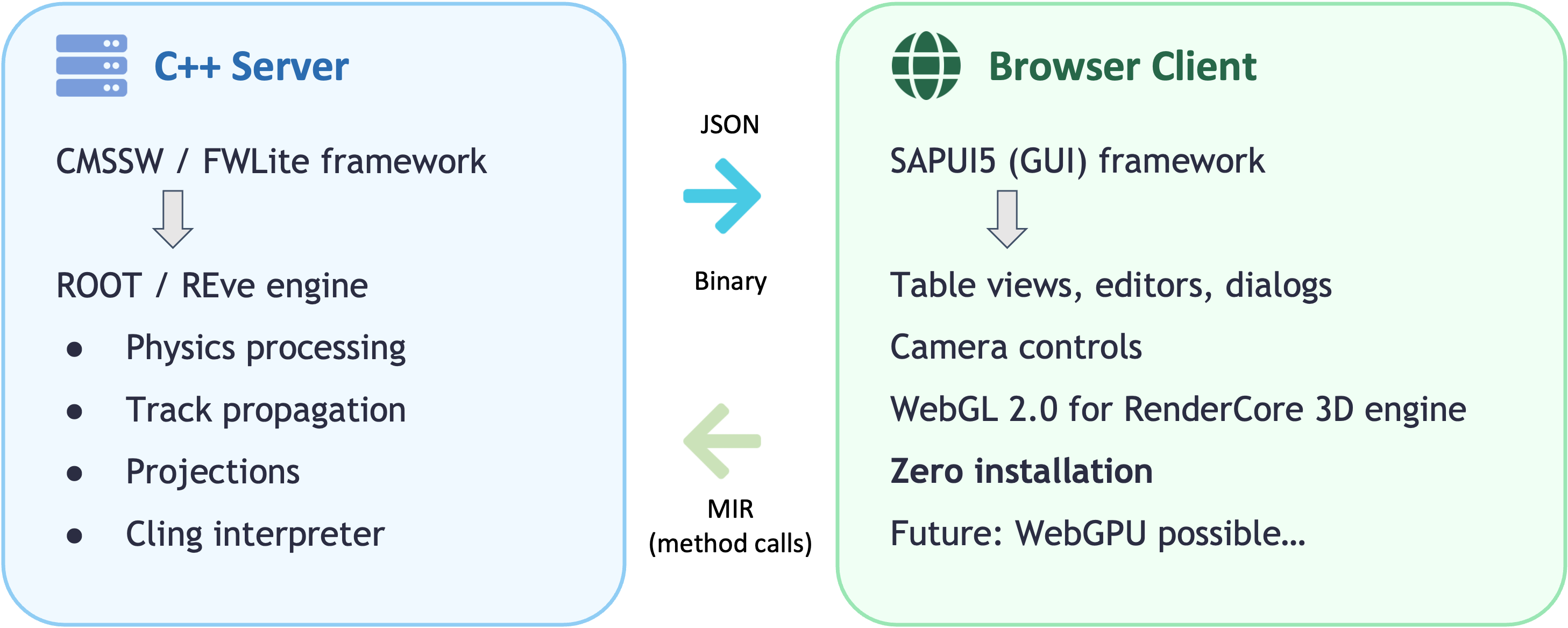}
\caption{The FireworksWeb server/client architecture. Physics processing stays in C\texttt{++} on the server; the browser handles presentation and interaction. Scene data reach the client as JSON and binary buffers; user actions return as Method Invocation Requests (MIR).}
\label{fig:arch}
\end{figure}

FireworksWeb uses a server/client design in which the physics-aware processing stays in C\texttt{++} on the server and the browser handles presentation and interaction (Fig.~\ref{fig:arch}). The server is built on the CMSSW/FWLite framework and the REve engine. It reads events, extracts and propagates tracks through the magnetic field, computes the RPhi and Rho-Z projections, and embeds the Cling C\texttt{++} interpreter, so that analysis macros and configuration can be evaluated at run time. It then turns the visual
representation of an event into REve elements and serializes them.
 
The client is a single-page application built on the SAPUI5 framework. It provides the table views, editors, and dialogs, the camera controls, and the WebGL~2.0 surface on which RenderCore draws. It needs no installation beyond a modern browser.
 
Communication runs over a WebSocket connection. From server to client, the scene and its updates are sent as JSON for structure and metadata, with binary buffers for the bulk numerical data (vertex positions, colours, indices). From client to server, user actions are sent as Method Invocation Requests (MIR): the client asks the server to invoke a named method on a given element, for example to change a collection filter or select a camera, and the server replies with the
resulting scene change. One authoritative copy of the event stays on the server, while the browser stays interactive. The same server can serve several clients, which is what the multi-user and remote-monitoring cases rely on. The production instance is deployed at
\href{https://fireworks.cern.ch}{fireworks.cern.ch}. A WebGPU rendering backend is a possible future extension of the client.

The unit of communication is the REve element. On the server an event is a tree of \code{REveElement} objects (collections, physics objects, geometry nodes, viewers, and cameras), each of which serializes the part of its state the client needs. When the scene changes, only the affected elements are re-serialized and pushed, so navigating between events or toggling a collection does not resend
the whole scene. The Cling interpreter is what keeps the configuration flexible: collection definitions, filter expressions, and data-loading macros are evaluated at run time, which the universal display of Sec.~\ref{sec:universal} later uses to read new inputs without recompilation. Because the browser holds no physics-aware code, it can only ask the server to invoke methods on existing
elements, and all interpretation of detector data happens on the server. Adding a new input format is therefore a server-side change alone, with nothing to modify in the client.

\section{Rendering and interaction}
\label{sec:display}
The 3D graphics and the interactive controls run in the browser, on top of the REve elements streamed by the server.

\subsection{RenderCore rendering}
\label{sec:rendercore}

\begin{figure}[t]
\centering
\includegraphics[width=0.65\linewidth]{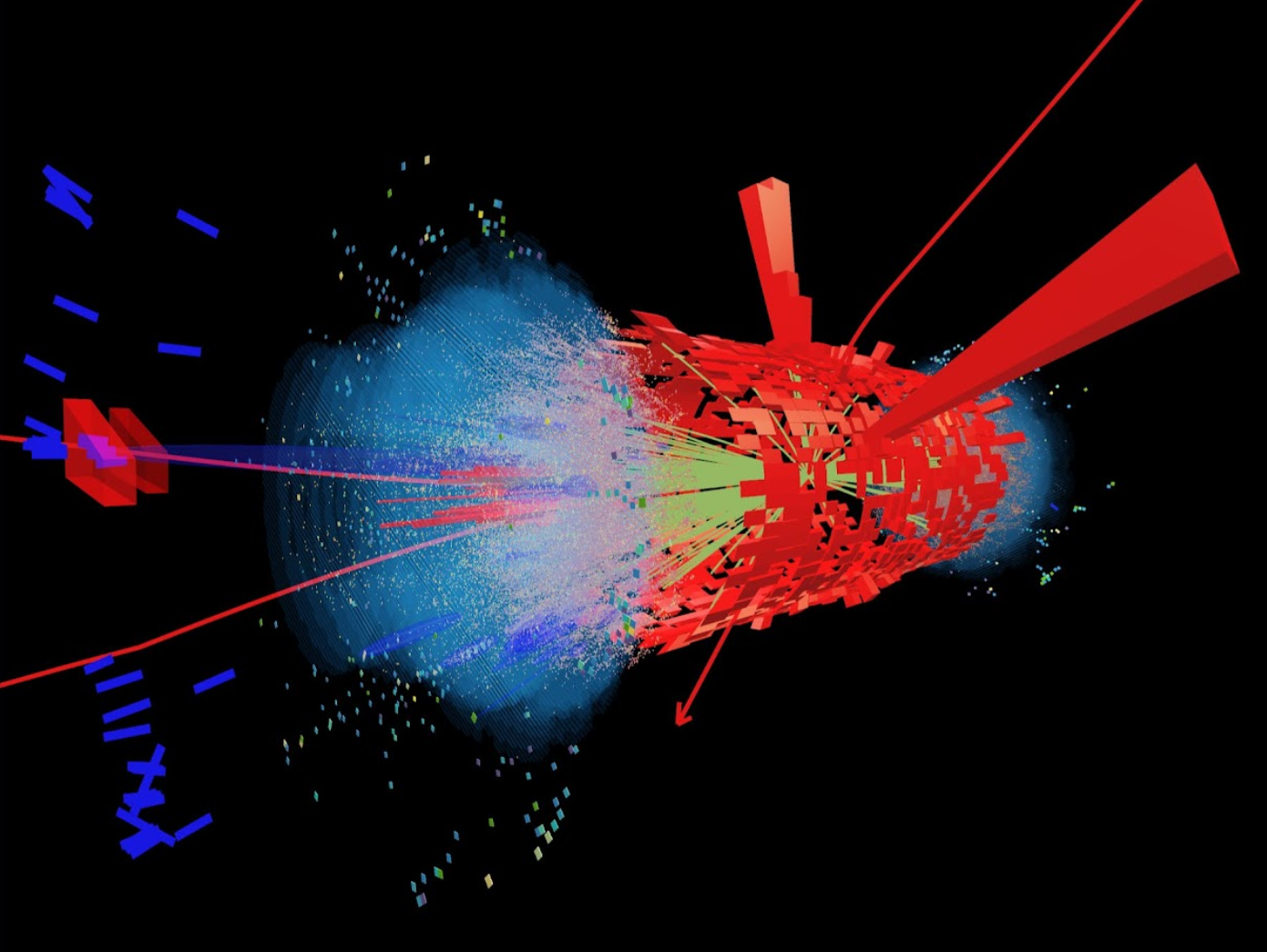}
\caption{HGCal visualization produced with the RenderCore engine, showing calorimeter deposits and propagated tracks. Instanced rendering and custom shaders scale to the $\sim$1.2~M hexagonal prisms of the HGCal. Reproduced from Ref.~\cite{rendercore} under CC-BY-4.0.}
\label{fig:rendercore}
\end{figure}

The 3D graphics are produced by RenderCore~\cite{rendercore}, a rendering engine developed for ROOT-Eve that replaced the earlier THREE.js~\cite{threejs} path. It gives explicit control over the WebGL pipeline, with custom shaders and instanced rendering, which is useful for the large, repetitive geometries of CMS subdetectors. The most demanding case is the High-Granularity Calorimeter (HGCal), whose display involves a number of order $1.2\times10^{6}$ hexagonal prisms; drawing these as instanced geometry keeps the scene interactive (Fig.~\ref{fig:rendercore}). RenderCore also resolves a pick to a single hit, tower, or track segment inside a large collection and maps it back to the physics object on the server. Because repeated primitives are drawn as instances with their per-object data in GPU buffers, adding more hits costs mainly data transfer, which keeps high-pileup events responsive. 

\subsection{Camera}
\label{sec:camera}
The camera is now a standalone, first-class element, independent of the viewer, so it can be shared, listed, and saved like any other part of the scene. Eleven predefined views are provided: three perspective cameras (XOZ, YOZ, XOY) and eight orthographic ones. The perspective views suit outreach and reading the overall topology of an event; the orthographic views give a clean transverse or longitudinal picture. The camera state, a base orientation and a transformation, is serialized by the server and applied by the client, and user changes are sent back so the server keeps the authoritative copy. The state therefore survives a reload or a shared session, which is what the view persistence and Open Data sharing rely on.

\subsection{Projection controls}
\label{sec:projection}
\begin{figure}[t]
\centering
\includegraphics[width=0.9\linewidth]{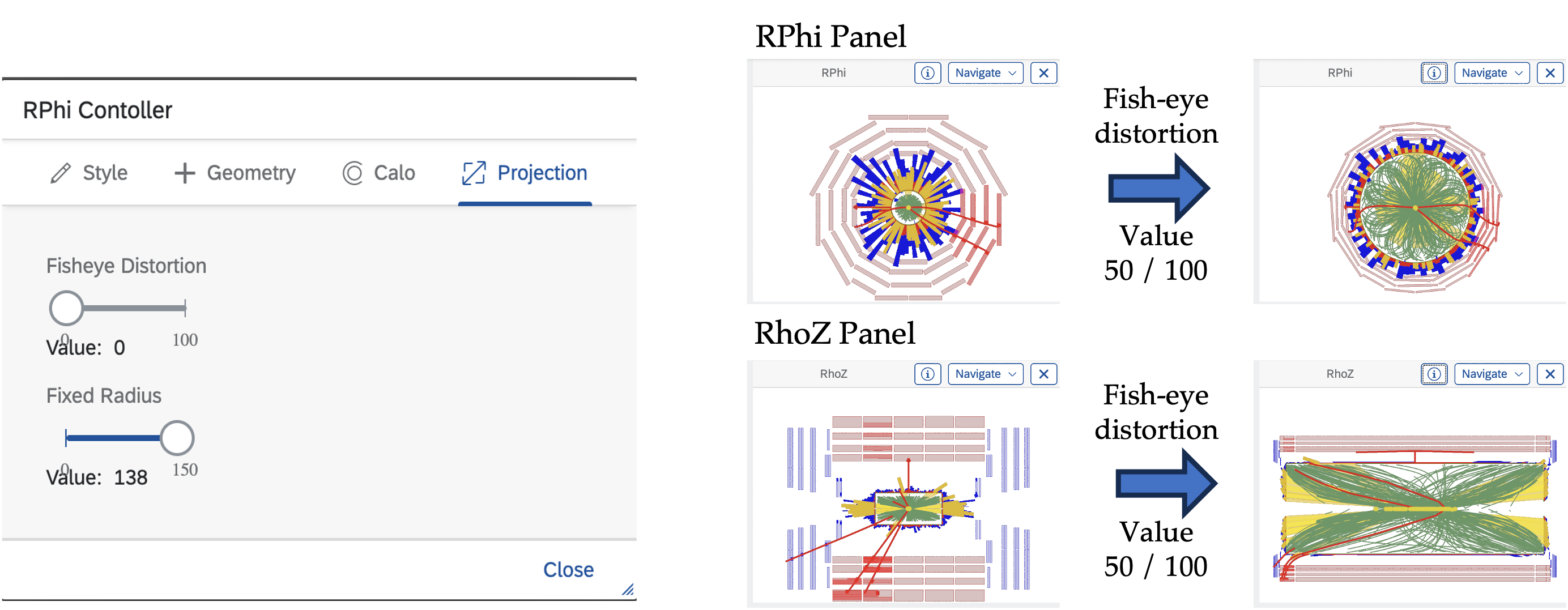}
\caption{Projection controls. The fish-eye strength and radius are set with step-input widgets (left); the RPhi (upper right) and Rho-Z (lower right) panels are shown before and after a distortion that pulls the outer detector layers inward while keeping the central region.}
\label{fig:projection}
\end{figure}

Events are read most easily in the transverse (RPhi) and longitudinal (Rho-Z) projections. The RPhi view, looking down the beam axis, shows azimuthal structure and the balance of transverse momentum; the Rho-Z view shows the forward and backward development of the event. In both, the detector spans radii that differ by more than an order of magnitude between the inner pixels and the outer muon chambers, so a linear drawing either crowds the centre or pushes the periphery off-screen. A tunable fish-eye distortion compresses the outer radii while leaving the centre, so the whole detector fits in one view (Fig.~\ref{fig:projection}); the strength and radius are adjustable, with defaults matched to the legacy Fireworks display. Separate compression for the muon and calorimeter regions is planned.

\subsection{View persistence and layout}
\label{sec:views}
Per-view settings (background, axes, layer ordering, and the camera) can be saved to a small configuration file and restored from the command line or from a URL, so a prepared view can be shared as a link. The file records how an event is shown, not the event data, so it is small and can be reapplied to another event of the same type. The drawing order of overlapping collections can also be set, so that, for example, tracks appear in front of jets.

\section{Open Data Access Pipeline}
\label{sec:pipeline}
\begin{figure}[t]
\centering
\includegraphics[width=0.7\linewidth]{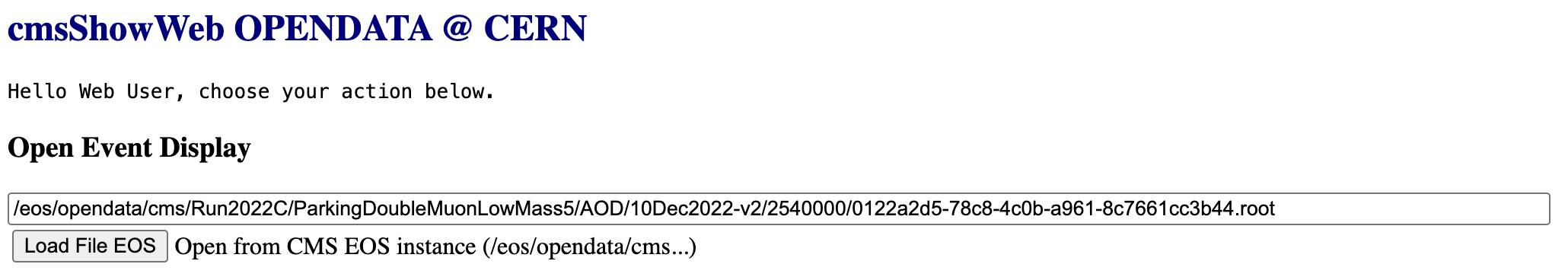}
\caption{The public low-level gateway at \texttt{revetor.pl}: a user opens a CMS
Open Data event by giving its EOS path directly, or picks a random sample.}
\label{fig:lowlevel}
\end{figure}

\begin{figure*}[t]
\centering
\includegraphics[width=0.85\textwidth]{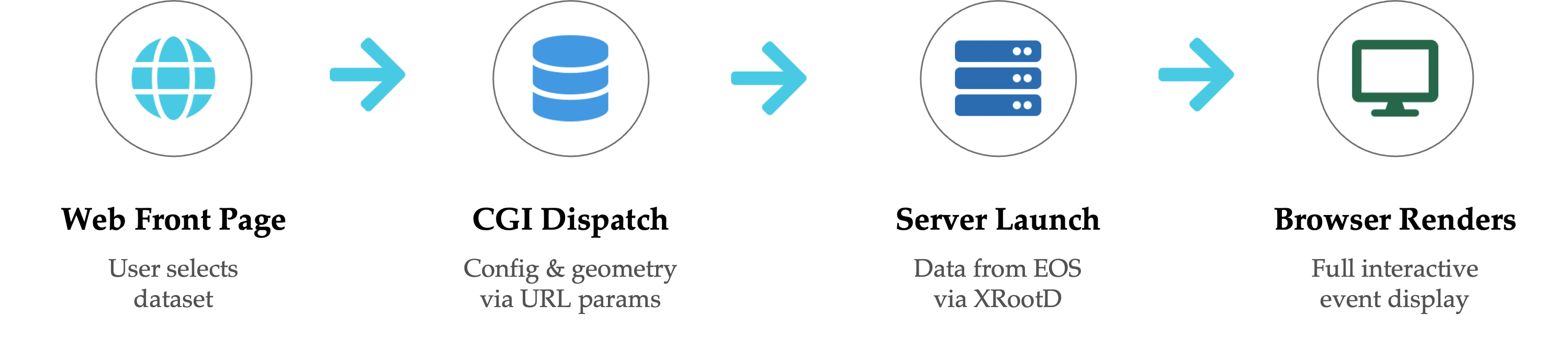}
\caption{The Open Data access pipeline. The user selects a dataset on the web front page; a CGI layer passes configuration and geometry as URL parameters; a per-user server process is launched and reads data from EOS over XRootD; and the browser renders the interactive display. No local installation or data download is required.}
\label{fig:pipeline}
\end{figure*}

The Open Data extension connects these components into a pipeline that takes a public user from a dataset choice to an interactive display without any local software (Fig.~\ref{fig:pipeline}). The entry point is a web front page on which the user selects a dataset or one of several predefined samples. The selection is passed through a CGI layer that hands the configuration and geometry to the server as URL parameters, which is also where a saved \code{.fwc} configuration or a custom geometry can be supplied. The server is then launched and reads the event data from CERN's EOS storage over the XRootD protocol, so the large Open Data samples are not downloaded by the user. The browser then renders the interactive display. Besides the predefined samples, a public low-level \href{https://fireworks-open.cern.ch/cmsShowWeb/revetor.pl}{gateway} lets a user open any CMS Open Data file on EOS directly by giving its path (Fig.~\ref{fig:lowlevel}).

A local forker service spawns a per-user server process that reads the requested file from EOS and streams the scene to that user's browser. The launch request carries the input file and optional configuration and geometry, which is where a saved view or a custom geometry is injected. Each user runs in an independent process with its own working area, so many simultaneous visitors, during a masterclass or an open day, do not interfere with one another. The browser never receives raw detector files and never runs physics-aware code. 

\section{Universal Event Display and VSDNano}
\label{sec:universal}
A direction that grew out of the Open Data work is to generalize the event display. So far it has read CMS data through Event Data Model files of CMSSW used by the AOD and MiniAOD tiers; by driving it instead from a compact, experiment-neutral event description, the same display can read the flat NanoAOD ntuple with plain ROOT, and ultimately data from other experiments. The universal display draws a small set of standard, high-level objects against a detector geometry. The event is filled by a short user routine, and a custom detector geometry can be supplied separately. The standard object set is small but physics-meaningful (Table~\ref{tab:objects}). Everything builds on a common candidate that carries momentum, position, charge, and an optional PDG id; jets add a cone radius and a hadronic fraction, muons a global-muon flag, and calorimeter towers and missing transverse energy their own quantities, while vertices and hits are space points and segments carry a position and a direction. Figure~\ref{fig:universal} shows this display running with generic objects. This part is still under development.
 
\begin{table}[t]
\centering
\caption{Standard objects currently defined in the VSD format and the properties used to draw each. The jet, muon, calorimeter tower, missing-$E_\mathrm{T}$ and segment types specialize the common candidate.}
\label{tab:objects}
\small
\begin{tabular}{@{}p{0.33\linewidth}p{0.59\linewidth}@{}}
\toprule
\textbf{Object} & \textbf{Key properties} \\
\midrule
Candidate & momentum ($p_\mathrm{T}$, $\eta$, $\phi$), position, charge, PDG id \\
Jet & momentum, cone radius, hadronic fraction \\
Muon & momentum, charge, global-muon flag \\
Calo tower & momentum ($p_\mathrm{T}$, $\eta$, $\phi$) \\
Missing $E_\mathrm{T}$ & $p_\mathrm{T}$, $\phi$, scalar $\sum E_\mathrm{T}$ \\
Vertex & 3D position with covariance \\
Hit & 3D position (space point) \\
Segment & position and direction (slopes) \\
\bottomrule
\end{tabular}
\end{table}
 
The format behind this is VSD (Visual Summary Data), a compact, experiment-neutral description of the objects an event display needs; historically the same idea has also been called a universal data format. VSDNano~\cite{vsdnano} is a small toolkit that writes a plain ROOT ntuple into a VSD file, which FireworksWeb reads directly. The workflow is light: build the VSD dictionaries, run a short Python script that fills the standard objects, and open the result in ROOT. It runs on the LCG software stack and needs neither CMSSW nor a compilation step for the mapping.
 
A VSD file holds a small set of object types: candidates (each with a PDG id shown in the table view), jets, muons, tracks, and vertices, and, for detector-level detail, hits (drawn as point sets) and track segments (drawn as polylines). Generator particles are handled in the same way. For NanoAOD the mapping is fixed, because the branch names are standard: muons are built from their momentum and charge with the trajectory reference point taken from the transverse and longitudinal impact parameters, and jets from their momentum.
 
Custom branches can be added on top of the standard ones. As an example we prepared a $B^{0}\!\to J/\psi\,K^{0}_{S}$ sample from NanoAOD. The $B^{0}$ and $J/\psi$ vertices are taken from the reconstruction, the $K^{0}_{S}$ vertex is computed from the $B^{0}$ vertex and the $K^{0}_{S}$ decay length along its flight direction, and the two pions and two muons are drawn with the three vertices (Fig.~\ref{fig:universal}). Primary vertices, generator particles, and trigger objects can be shown as well, which helps with background studies in $B$ physics. Because the display reads VSD rather than a CMSSW data format, a group with a custom ntuple can get an interactive, browser-based view of their data by writing a short script, reusing everything described above.
 
Two smaller fixes improve the projected views of the VSD display: looping low-momentum tracks are truncated after half a turn so they do not clutter the picture, and in the Rho-Z projection the projection plane is rotated so that a track crossing the beam axis is drawn as one piece rather than split in two, which had been a visible artifact for pion tracks.
 
\begin{figure}[t]
\centering
\includegraphics[width=0.8\linewidth]{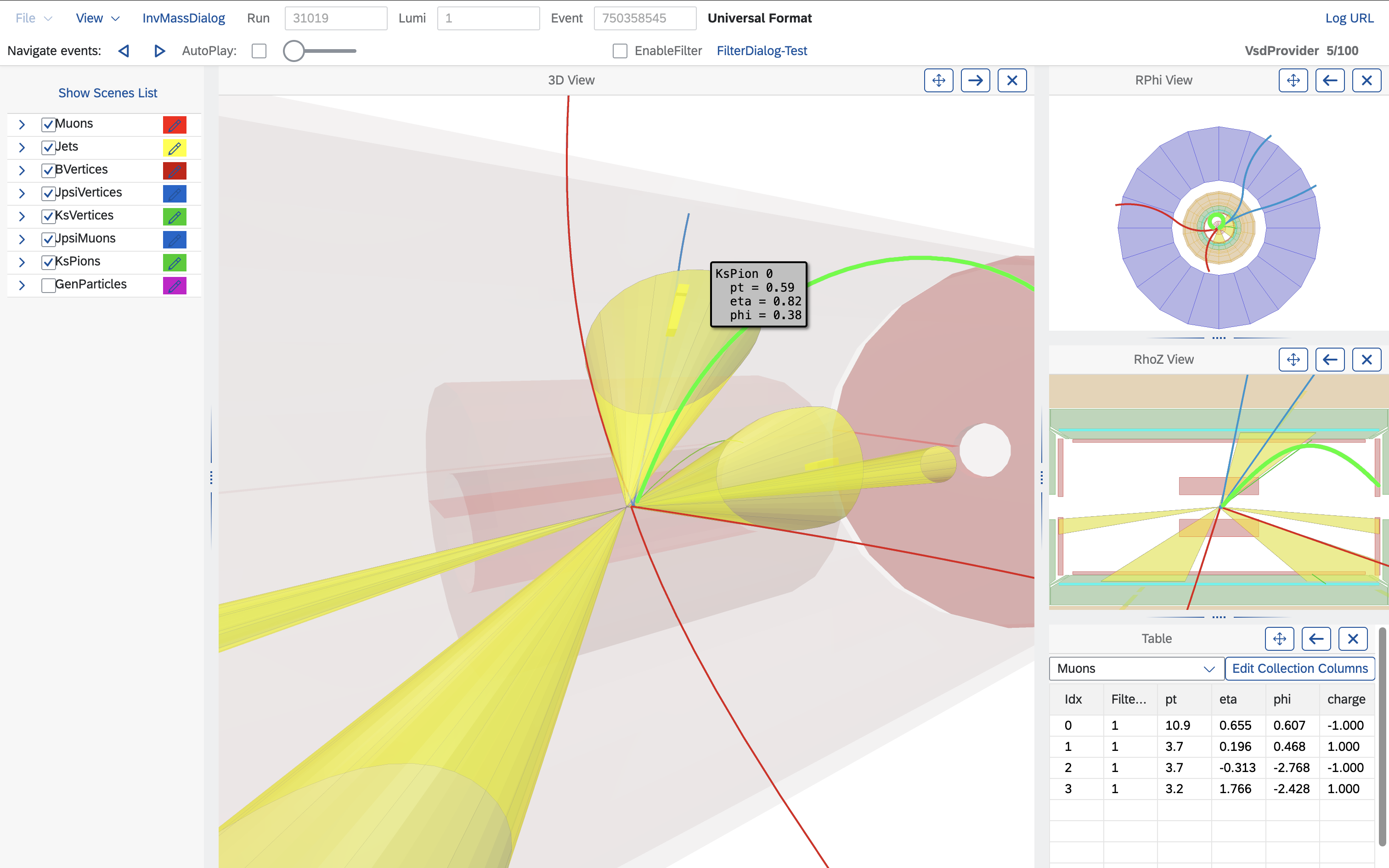}
\caption{An event from the $B^{0}\!\to J/\psi\,K^{0}_{S}$ NanoAOD sample in the universal display. the two muons from the $J/\psi$ and the two pions from the $K^{0}_{S}$ are drawn together with the reconstructed $B^{0}$, $J/\psi$, and $K^{0}_{S}$ vertices. The same browser interface serves full CMS events, with no CMSSW dependency.}
\label{fig:universal}
\label{fig:universal}
\end{figure}

\section{Towards the HL-LHC}
\label{sec:hllhc}
Further development follows the needs of CMS for the High-Luminosity LHC. One direction is support for analysis: closer integration with the Python ecosystem and a way to export user-defined objects for display, for which VSDNano (Sec.~\ref{sec:universal}) is the starting point. A second is use as a tool for debugging reconstruction, with a first focus on tracking and HGCal, where inspecting individual hits and their association to reconstructed objects is valuable. A third is the ongoing commissioning of the CMS detector, trigger, and software, where a remote, control-room-tested display is already in use.

\section{Conclusion}
\label{sec:summary}
FireworksWeb is the production CMS event display, and this work extends it to the CMS Open Data as a prototype. It is deployed at fireworks.cern.ch and has been used in the CMS control room, and it gives browser-based access to real CMS collision data with nothing to install. The components described here are a standalone camera with eleven views and matched server and client state, interactive projection controls, saving and restoring of per-view configurations, and a CGI pipeline that reads Open Data from EOS over XRootD. VSDNano writes plain ntuples into the VSD format, so the same display can show data without a CMSSW dependency, which is also the basis of the universal event display.
 
Planned work includes the VSD universal data format, a WebGPU backend through RenderCore, further work on HGCal visualization, and integration with the CERN Open Data portal.

\bibliography{references}

\end{document}